\documentstyle[12pt]{article}
\begin{document}
\begin{center}
{ \large\bf Faraday's law in the presence of magnetic monopoles\\}
\vskip 1cm

{M. Nowakowski and N. G. Kelkar\\}
Departamento de Fisica, Universidad de los Andes,
Cra.1 No.18A-10, Santafe de Bogota, Colombia
\end{center}
\begin{center}
\end{center}
\begin{abstract}

We show that if we consider the full statement of
Faraday's law for a closed physical circuit,
the standard Maxwell's equations in the presence of electric and
magnetic charges have to include in their integral
form a mixed term of the form $\rho_m {\bf v}_e^{\perp}$ where $\rho_m$
is the magnetic charge density and ${\bf v}_e^{\perp}$ the perpendicular
component of the velocity ${\bf v}_e$ of the electric charge. 

\end{abstract}
\noindent

PACS: 03.50.-z, 03.50.De, 14.80.Hv
\vskip 1cm
\noindent
Maxwell's electrodynamics with its model role for a fundamental
theory \cite{raif} and its numerous applications is one of the most 
successful
theories in physics \cite{jackson, books}.
The equivalence between Maxwell's laws in integral and differential form
is evident in the derivation of the latter from the former and
 manifest in the
claim that the Maxwell's equations in differential form together with
the Lorentz force encompass the whole of electromagnetism.
The integral form of Faraday's law plays a special role in this
context. Let us consider a real closed physical circuit with moving 
boundaries. If we wish to include the motional induced electromotive 
force (emf) in the Faraday's law, the boundary of the surface integral
in this law becomes time dependent. 
In such a situation the velocity ${\bf v}_{pull}$ which is controlled
externally can depend
both on position and time, i.e., ${\bf v}_{pull} = {\bf v}_{pull}(t, {\bf 
r})$.
It is not difficult to start with the Faraday's law with a time dependent
boundary which encompasses the motional induced emf to derive the 
third Maxwell's equation in differential form. However, this procedure 
requires a mathematical identity for differentiating surface integrals
in which the divergence of the magnetic field will appear. 
One of the reasons why the above mentioned general equivalence of the
differential and integral form of Faraday's law
(including motional emf) can be
proved, is 
the validity of the second Maxwell's equation, 
${\bf \nabla} \cdot {\bf B} = 0$, i.e., 
the non-existence of magnetic monopoles. 
One would of course expect that motional emf is also included in
the Maxwell's equations valid for electric and speculative magnetic
charges \cite{dirac,reviews,monopole, searches} 
as the former is an experimental fact.
However, the presence of magnetic charges, i.e. 
$\nabla \cdot {\bf B} \ne0$, 
requires a reanalysis which leads to a novel result as shown below.

We first touch upon the standard case of Maxwell's electrodynamics to
clarify certain tacit features about the explicit inclusion
of motional emf into Maxwell's equations. Although not always a
common practice in literature, it is in general known that
to include the motional emf case in the integral form of Faraday's law, 
the latter has to be written in the following form 
\cite{jackson, frankel, lea} :
\begin{eqnarray}\label{faraday}
&&\frac{1}{e}\oint_C {\bf F}_{{\rm Lorentz}, \,e}
\cdot d{\bf l} 
 = -\frac{d}{dt}\Phi_B=-\frac{d}{dt}\int_S {\bf B}\cdot d{\bf A}\,, 
\\ \nonumber
&&\oint_C ({\bf E} + {\bf v}_e
\times {\bf B}) \cdot d{\bf l}=\oint_C ({\bf E}
+ {\bf v}_e^{\perp} \times {\bf B}) \cdot d{\bf l} =
-\frac{d}{dt}\int_S {\bf B}\cdot d{\bf A}
\end{eqnarray}
where ${\bf F}_{{\rm Lorentz},\,e} =e( {\bf E}
+ {\bf v}_e \times {\bf B})$, is the usual Lorentz force on an
electric charge $e$ and $C$ denotes the closed path around the surface $S$.
The velocity (${\bf v}_e$) of the {\em electric} 
charge,
$e$, has been split according to ${\bf v}_e
={\bf v}_e^{\perp} + {\bf v}_e^{||}$ with ${\bf v}_e^{||}$ parallel to
$d{\bf l}$ and ${\bf v}^{\perp}$ the perpendicular component which, in most
applications of the electromotive force, is the velocity
${\bf v}^{pull}$ with which the rod is pulled. 
The second line in (\ref{faraday}) is to indicate that only the term in 
${\bf v}_e^{\perp}$ survives.
By writing equation (\ref{faraday}) we explicitly allow for
variable boundaries. It is of some importance to stress that the
velocity ${\bf v}_e$ can be, in general, a velocity flow vector field
i.e. dependent on time and position vector (${\bf v}_e={\bf v}_e(t, {\bf 
r})$).
This is so because ${\bf v}^{pull}$ is controlled by an external agent.
The mathematical identity to handle such a general case is \cite{frankel}
\begin{equation}\label{frankel}
{d \over dt} \int_S {\bf G} \cdot  d{\bf A} = \int_S \biggl [ {\partial
{\bf G} \over \partial t } + (\nabla \cdot {\bf G}) {\bf v} - \nabla \times
({\bf v} \times {\bf G}) \biggr ]\cdot d{\bf A}
\end{equation}
for an arbitrary vector field ${\bf G}$ \cite{jackson2}.
Specializing to ${\bf G}={\bf B}$ together
with $\nabla \cdot {\bf B}=0$ and applying Stokes's theorem to 
(\ref{faraday}),
we get from (\ref{faraday}) the third Maxwell's equation
\begin{equation} \label{3maxwell}
\nabla \times {\bf E} = -\frac{\partial {\bf B}}{\partial t}\,.
\end{equation}
If in the above discussion
we are talking about electron's velocity ${\bf v}_e$, we then have
explicitly in mind  that the path $C$ is a closed physical circuit
in which the electrons move. Technically, we can also consider the case 
where
$C$ is a geometrical path moving in space with velocity ${\bf v}$
relative to the observer \cite{jackson}.
Then in (\ref{faraday}) we need only to replace
${\bf v}_e$ by ${\bf v}$ without changing the meaning of Faraday's equation.
Of course, ${\bf v}$ is then an `artificial' construct and should not enter 
e.g.
the differential relation between fields which is the case in 
(\ref{3maxwell}).
Similarly the velocity ${\bf v}$ of the imaginary loop is irrelevant in
interpreting the induced emf as ${\cal E}_{emf}=-d\Phi_B/dt$.

From the mathematical point of view, it might be possible to derive
many integral identities from (\ref{3maxwell}). The point is of course that
the integral law (\ref{faraday}) has a direct physical meaning: the left 
hand
side is the measured induced emf, ${\cal E}_{emf}$.
For instance, in a conducting
material, the induced current would be $I= {\cal E}_{emf}/R$, with $R$ the
resistance of the material.

The equivalence between (\ref{faraday}) and (\ref{3maxwell}) is established
by integrating the differential law over a surface $S$ and using
Stokes's theorem in the form $\int_S (\nabla \times {\bf E})\cdot d{\bf A}
=\oint_{C}{\bf E} \cdot d{\bf l}$ and $\int_S [\nabla \times (
{\bf v}_e \times {\bf B})]\cdot d {\bf A}=\oint_{C}
({\bf v}_e \times {\bf B})\cdot d{\bf l}$. Again the absence of magnetic
monopoles is one of the main assumptions if we want to recover 
(\ref{faraday})
from (\ref{3maxwell}). In view of this result it is legitimate to put
forth the question of how can one ensure the equivalence of the
differential and integral laws once the magnetic monopoles are introduced
into electromagnetism.
Of course, this question should also take into account the
full statement of Faraday's law which mathematically manifests itself
in the {\em total} time derivative of the magnetic flux $\Phi_B$. With
standard notation for charge and current densities, the following
set of Maxwell's equations
\begin{eqnarray} \label{maxwellmag}
\nabla \cdot {\bf E} &=& {\rho_e \over \epsilon_0} \nonumber \\
\nabla \cdot {\bf B} &=& \mu_0 \rho_m \nonumber \\
\nabla \times {\bf E} &=& - {\partial {\bf B} \over \partial t}
-\mu_0 \, {\bf J}_m \nonumber \\
\nabla \times {\bf B} &=& \epsilon_0 \mu_0 {\partial {\bf E} \over \partial 
t}
+ \mu_0 {\bf J}_e
\end{eqnarray}
is believed to govern the whole of electromagnetism, once electric ($e$) and
magnetic ($g$) charges \cite{monopole} are introduced.
Going back from the third Maxwell's equation in (\ref{maxwellmag}), like 
from
(\ref{3maxwell}) to (\ref{faraday}), one would intuitively expect to get
the integral law, $\frac{1}{e}\oint_{C}{\bf F}_{{\rm Lorentz}, \,e}
\cdot d{\bf l}=-\frac{d}{dt}\Phi_B-\mu_0\int_S {\bf J}_m \cdot d{\bf A}$.
This is however not the case.
The correct integral law follows from $\nabla \cdot {\bf B} =\mu_0 \rho_m$
and the mathematical identity (\ref{frankel}). In the presence of magnetic
monopoles, the induced emf, ${\cal E}_{emf}$ is,
\begin{equation}\label{integralnew}
\frac{1}{e}\oint_{C}{\bf F}_{{\rm Lorentz}, \,e}\cdot d{\bf l}
=-\frac{d}{dt}\Phi_B -\mu_0 \int_S {\bf J}_m \cdot d{\bf A} +
\mu_0 \int_S \mbox{\boldmath $\iota$}_{m,\,e} \cdot d{\bf A}
\end{equation}
with an unusual mixed term
\begin{equation} \label{mixedterm}
\mbox{\boldmath $\iota$}_{m,\, e} = \rho_m {\bf v}_e^{\perp}\, , 
\end{equation}
We remind the reader that ${\bf v}_e^{\perp}$ is the perpendicular
component of the electron's velocity ${\bf v}_e$ with respect to the loop (obviously
it is also the velocity of the loop). 
Note that we insist here on the interpretation of $C$ as a closed
physical circuit. Otherwise, in the case of $C$ being an imaginary
loop, the measurable quantity ${\cal E}_{emf}$ would depend on an unphysical
variable ${\bf v}$.  By itself it is a remarkable fact that in the
presence of magnetic monopoles, it does not make much sense to allow
$C$ to be a moving imaginary loop as can be the case in
(\ref{faraday}). Indeed, (\ref{integralnew}) is meant as a mathematical
expression of Faraday's experiment in a circuit with moving boundaries. This
is an interesting difference as compared to the standard case.
The difference is technical and of course, in a physical situation, 
where the charges move solely according to the Lorentz force (i.e 
${\bf v_e} = {\bf v_e}^{||}$) the term with 
$\mbox{\boldmath $\iota$}_{m,\, e}$ in (\ref{integralnew}) vanishes as 
it should be.

Two points are worth mentioning regarding this new term. Firstly,
$\mbox{\boldmath $\iota$}_{m,\,e}$ is in general not a current 
density like
${\bf J}_e=\rho_e {\bf v}_e$ and ${\bf J}_m=\rho_m{\bf v}_m$ where
${\bf v}_m$ is the velocity of the magnetic charges unless we are
talking about dyons whose case will be discussed below.
The mixing of a magnetic ($\rho_m$)
property with an electric one (${\bf v}_e$) is unusual, but a
straightforward consequence if we insist on including the motional emf into
the integral law with monopoles. 
The fact that the mixed term in (\ref{mixedterm}) is not a current density gets also
reflected in the difference between the microscopic definitions
of ${\bf J}_e$ and $\mbox{\boldmath $\iota$}_{m,e}$. In the microscopic version
of ${\bf J}_e$, namely, 
\begin{equation} \label{Je}
{\bf J}_e({\bf r}, t)=\sum_a e_a{\bf v}_a(t)\delta^{(3)}({\bf r}-{\bf r}_a(t))\, ,
\end{equation}
we sum all properties of electrons over all electrons. The microscopic
analog of $\rho_m({\bf r}, t)$ is $\sum_ag_a\delta^{(3)}({\bf r} -{\bf r}_a(t))$ 
where $g_a$ is the individual magnetic charge. Hence we can write
\begin{equation} \label{micro}
\mbox{\boldmath $\iota$}_{m, e}({\bf r}, t) = \sum_a g_a\delta^{(3)}({\bf r} -{\bf r}_a(t)){\bf v}_e^{\perp}({\bf r}, t)
\end{equation}
There is no sum in connection with the velocity and it is not necessary to `discretize' the velocity
as the latter is the velocity of the loop and hence the same for each electron in the circuit at the
position ${\bf r}$ and at a time $t$. The best way to visualize it is to consider a constant loop velocity
of a straight piece of loop
(like in the standard motional emf experiment)
in which case all electrons in this part of the circuit will have the same
perpendicular velocity. 

The second remark concerns the
significance of such a term. It is beyond the scope of this letter to
give a full account of this issue. We can, however, draw the 
reader's attention to the
fact that although $\int_S \mbox{\boldmath $\iota$}_{m,\,e}\cdot d{\bf A}$
cancels against $-\int_S {\bf v}_e^{\perp}(\nabla \cdot {\bf B})\cdot d{\bf A}$
in $-\frac{d\Phi_B}{dt}$ (see (\ref{frankel})),
this does not mean that such a term has no
significance. The contrary is the case. Either for a given magnetic field
${\bf B}$ we calculate the magnetic flux and its time derivative directly,
which leaves us with the mixed term $\int_S \mbox{\boldmath $\iota$}_{m,\,e}
\cdot d{\bf A}$ or by using the identity
\begin{equation}\label{identity}
\nabla \times ({\bf k} \times {\bf G}) = {\bf k} (\nabla \cdot {\bf G} )
- {\bf G} (\nabla \cdot {\bf k} ) + ({\bf G} \cdot \nabla ) {\bf k} -
( {\bf k} \cdot \nabla ) {\bf G}
\end{equation}
we can convince ourselves that
\begin{equation} \label{effect}
-\frac{d\Phi_B}{dt} +\mu_0 \int_S \mbox{\boldmath $\iota$}_{m,\,e}
\cdot d{\bf A} =\int_S\left[ -\frac{\partial {\bf B}}{\partial t} + \mu_0
\mbox{\boldmath $\iota$}_{m,\,e} - ({\bf v}_e^{\perp} \cdot \nabla) {\bf B}
\right ]\cdot d{\bf A}
\end{equation}
where for simplicity we assumed that the velocity depends only on time
$t$. In simple words, the mixed term
$\mbox{\boldmath $\iota$}_{m,\,e}$ is also contained in
$\nabla \times ({\bf v}_e^{\perp} \times {\bf B})$.

Suppose that we perform the standard motional emf experiment known from
textbooks, but now in the presence of very heavy (static) magnetic 
monopoles.
This assumption entitles us to put ${\bf J}_m$ approximately to zero.
Then naively one might suspect that $-\frac{d\Phi_B}{dt}$ is the only
contribution to the emf. The above results show, however, that
$\rho_m$ and ${\bf v}_e$ combined in $\mbox{\boldmath $\iota$}_{m,\,e}$
give an additional contribution which is unconventional and worth pointing out.

One of the motivation to introduce magnetic monopoles was to obtain
the symmetric form of the Maxwell's equations.
One can establish a completely symmetric
form of Maxwell's equations for magnetic monopoles also in the integral 
form.
Especially, we have in mind a symmetry between the new Faraday's law
(\ref{integralnew}) and the Ampere-Maxwell law. If we make use
of the force ${\bf F}_{{\rm Lorentz}, \,g}$
acting on magnetic charges $g$, i.e.
\begin{equation} \label{magforce}
\frac{d {\bf p}_m}{dt} =g \left[\frac{{\bf B}}{\mu_0} -\epsilon_0
{\bf v}_m \times {\bf E}\right]\,,
\end{equation}
we can write for the integral Ampere-Maxwell law
\begin{equation}\label{integralnew2}
\frac{1}{g}\oint_{C}{\bf F}_{{\rm Lorentz}, \,g}\cdot d{\bf l}
=\epsilon_0\frac{d}{dt}\Phi_E + \int_S {\bf J}_e \cdot d{\bf A} -
\int_S \mbox{\boldmath $\iota$}_{e,\,m} \cdot d{\bf A}
\end{equation}
with
\begin{eqnarray} \label{definitions}
\Phi_E &=&\int_S {\bf E}\cdot d{\bf A} \nonumber \\
\mbox{\boldmath $\iota$}_{e,\,m}&=&\rho_e {\bf v}_m^{\perp} \, .
\end{eqnarray}
Then (\ref{integralnew2}) is in perfect symmetrical analogy to
(\ref{integralnew}) and equivalent to the fourth differential equation
in (\ref{maxwellmag}). Of course, in the presence
of magnetic monopoles, (\ref{integralnew}) is necessitated by the
experimental fact of motional emf whereas such support is lacking
for (\ref{integralnew2}). 
We base (\ref{integralnew2}) on the fact that the Maxwell equations 
are symmetric with regard to electric and magnetic charges and so 
should be any conclusion drawn from them, unless this symmetry
is broken from outside.
The real relevance of (\ref{integralnew2}) is indeed
in interpreting the left hand side of it as an induced magnetomotive
force (mmf) (in analogy to emf), resulting in an induced magnetic
field in an experiment analogous to the motional emf one.
Hence, similar to (\ref{integralnew}), $ \mbox{\boldmath$\iota$}_{e,\,m}$
appears only if the magnetic field is induced motionally. 
Interestingly, for the case of dyons for which we have ${\bf v}_e=
{\bf v}_m$, such a symmetry of Maxwell's equations in the integral form
would result into a very simple form
\begin{equation} \label{dyons}
\frac{1}{g_i} \oint_{C} {\bf F}_{{\rm Lorentz},\,g_i}
\cdot d{\bf l}=-\epsilon_i \frac{d}{dt} \Phi_i
\end{equation}
where $i=E,\, B$, $\epsilon_E=-\epsilon_0$, $\epsilon_B=1$ and
$g_E=e$, $g_B=g$. Indeed, the surface integrals over the current densities 
${\bf J}_e$ and ${\bf J}_m$, then emerge automatically
from $-\frac{d\Phi_E}{dt}$ and $-\frac{d\Phi_B}{dt}$, respectively.

From a purely mathematical point of view, there is 
at least in principle, yet another possibility to
reconcile the integral laws with the differential ones when both electric
and magnetic charges are present. We can also incorporate the mixed term(s)
directly into the differential law(s). The modified third Maxwell's
equation in (\ref{maxwellmag}) could read
\begin{equation} \label{newmaxwell}
\nabla \times {\bf E} = - {\partial {\bf B} \over \partial t}
-\mu_0 \, {\bf J}_m \nonumber  -\mu_0
\mbox{\boldmath $\iota$}_{m,\,e}
\end{equation}
and eventually guided by symmetry principles as before,
the fourth one could read
\begin{equation} \label{newmaxwell2}
\nabla \times {\bf B} = \epsilon_0 \mu_0 {\partial {\bf E} \over \partial t}
+ \mu_0 {\bf J}_e +\mu_0 \mbox{\boldmath $\iota$}_{e,\,m}\, .
\end{equation}
In such a situation the corresponding equivalent integral laws are
(\ref{integralnew}) and (\ref{integralnew2}), but {\em without}
the mixed terms on the right hand sides which explicitly demonstrates
that the physics of Faraday's law would be different from
(\ref{integralnew}) in this case.
With (\ref{newmaxwell}) and (\ref{newmaxwell2}) the equivalence between
differential and integral laws is again restored and the motional
manifestation of Faraday's law still included in the laws.
We note, however, that the second
possibility defined through equations (\ref{newmaxwell}) and
(\ref{newmaxwell2}) has some drawbacks. First of all the
continuity equations take now the form
\begin{eqnarray} \label{continuity}
\frac{\partial \rho_e}{\partial t} +\nabla \cdot \left({\bf J}_e
+\mbox{\boldmath $\iota$}_{e,\,m}\right)&=&0 \nonumber \\
\frac{\partial \rho_m}{\partial t} +\nabla \cdot \left({\bf J}_m
+\mbox{\boldmath $\iota$}_{m,\,e}\right)&=&0
\end{eqnarray}
which leave the global conservation of the electric and magnetic charges
untouched, but are difficult to interpret locally.
Secondly, if the usual dual transformations are supplemented
by the rule that
$\mbox{\boldmath $\iota$}_{m,\,e}$ transforms as ${\bf J}_m$ and
$\mbox{\boldmath $\iota$}_{e,\,m}$ as ${\bf J}_e$ we still find
the Maxwell's equations (\ref{newmaxwell}) and (\ref{newmaxwell2})
covariant under such a transformation, but it is
not possible to `transform away' $\rho_m$, ${\bf J}_m$,
$\mbox{\boldmath $\iota$}_{m,\,e}$ and $\mbox{\boldmath $\iota$}_{e,\,m}$ 
simultaneously.
From a physical point of view, the modifications (\ref{newmaxwell}) and
(\ref{newmaxwell2}) have two defects. As already stated, the continuity 
equations (\ref{continuity}) cannot be interpreted locally. For instance, 
the change in charge density with time is accounted by the term 
$\nabla \cdot {\bf J}_e$ and the remaining term in 
(\ref{continuity}) cannot be interpreted. Therefore, we refute 
(\ref{newmaxwell}) and (\ref{newmaxwell2}) on physical grounds and 
recognize (\ref{integralnew}) and (\ref{integralnew2}) 
as the correct integral laws which take the differential form as in 
(\ref{maxwellmag}).

In summary, we have shown that the general Faraday's law
(i.e. the one which encompasses also the electromotive case)
in the presence of magnetic monopoles,
requires the mixed term $\mbox{\boldmath $\iota$}_{m,\,e}$ as a source term
in the integral version of the Maxwell's equations.
To achieve fully symmetric equations in the integral
form requires also the introduction of the corresponding term
$\mbox{\boldmath $\iota$}_{e,\,m}$.
The choice to include such terms in the differential form results in a
continuity equation which locally seems to lack a simple interpretation.
The change in the magnetic charge density with time,
is still compensated by an outgoing flux, but the
latter contains the mixed term $\mbox{\boldmath $\iota$}_{m,\,e}$.
It is clear that such a mixed term in the integral form of
Faraday's law changes its physical content as discussed in the text.


\begin{thebibliography}{99}
\bibitem{jackson2}
The identity can also be found in \cite{jackson} where it is used for
velocity depending only on time.
\end{thebibliography}
\end{document}